\documentclass[prc,aps,nofootinbib,showkeys,showpacs,twocolumn]{revtex4}
\usepackage{epsfig}
\usepackage{graphicx}
\usepackage{amssymb}
\usepackage{color}
\begin{document}

\title{Polarization of the nuclear surface in deformed nuclei}

\author{Guillaume Scamps}
 \email{scamps@ganil.fr}
\affiliation{GANIL, CEA/DSM and CNRS/IN2P3, Bo\^ite Postale 55027, 14076 Caen Cedex, France}
\author{Denis Lacroix} \email{lacroix@ipno.in2p3.fr}
\affiliation{Institut de Physique Nucl\'eaire, IN2P3-CNRS, Universit\'e Paris-Sud, F-91406 Orsay Cedex, France}
\author{G.G.Adamian and N.V.Antonenko}
\affiliation{Joint Institute for Nuclear Research, 141980 Dubna, Russia}

\begin{abstract}
The density profiles of around 750 nuclei are analyzed using the Skyrme energy density
functional theory. Among them, more than 350 nuclei are found to be deformed. In addition to rather standard
properties of the density, we report
a non-trivial behavior of the nuclear diffuseness as the system becomes more and more
deformed. Besides the
geometric effects expected in rigid body, the diffuseness acquires a rather complex behavior
leading to a reduction of the diffuseness along the main axis of deformation simultaneously with an increase
of the diffuseness along the other axis. The possible isospin dependence of this polarization is studied.
This effect, that is systematically seen in medium- and heavy-nuclei, can affect the nuclear dynamical properties.
A quantitative example is given with the fusion barrier in the $^{40}$Ca+ $^{238}$U reaction.
\end{abstract}

\keywords{nuclear shapes,mean-field, pairing}
\pacs{21.10.Gv, 21.60.Jz, 25.60.Pj}

\maketitle

\section{Introduction}

Average central densities, nuclear radius of surface diffuseness are basic quantities
characterizing finite nuclear systems. Their precise study are expected to provide valuable
information leading to a global understanding of the nuclear many-body problem. To quote some of the
important information inferred from the nuclear density profile:
the almost constant central densities seen in nuclei gives a direct inside in the
incompressible nature of nuclear matter; extensive discussion are made currently
on the difference between proton-neutron nuclear radius to
study the symmetry energy \cite{Vin12,Che05,Hor01}.
Such considerations are usually restricted to spherical nuclei. Most of nuclei however
are found to be deformed in their ground states especially in the
medium- and heavy-mass regions \cite{Ben06,Del10}. The present work is an attempt to
identify some peculiar effects that might show up in deformed systems.

The energy density functional (EDF) based on Skyrme effective interaction is
a suitable tool for such a study. Indeed, following the same philosophy as the
density functional theory in electronic systems,
the nuclear energy density functional theory is optimized to provide an adequate
approximation for the total energy and the nuclear local density. The EDF is commonly used
nowadays to study the onset of deformation \cite{Ben03,Rin80}.
 Here, we make a large-scale analysis of the nuclear density profiles along the whole
nuclear chart.

A particular attention is made to study the diffuseness of the surface which have an important influence for the calculation of deformation \cite{Hag06,Mye98} as well as for the calculation of the fusion barrier \cite{Dut10}.
While the central density and radius can be understood with rather simple arguments,
properties of the nuclear surface diffuseness turns out to be more complicated than anticipated.

The article is organized as follows: in the next chapter, the methodology to characterize nuclear
shapes and density profiles is exposed. Section III provides a detailed analysis of the density profile
uncovering the quasi-systematic polarization of the nuclear surface. In section IV, an example of consequences
on nuclear dynamics is given.

\section{Methodology}

The nuclear density obtained with the {\sl EV8} code \cite{Bon05} was investigated over 749 even-even nuclei ranging from $Z=8$ to $Z=108$ whose experimental masses are known. The {\sl EV8} code solves the
HF+BCS in r-space with the Skyrme functional in the mean-field term and with a contact
interaction in the pairing channel. Here a surface type pairing interaction is used with the same strength as in the ref. \cite{Ber06}. 
In practice, the mean-field equations are only solved
in 1/8 of the space using specific symmetries.
The mesh size has been taken as $\Delta x =\Delta y = \Delta z  = 0.8$ fm and the total
size of the mesh are $2L_x = 2L_y = 2L_z = 28.8$ fm. We have checked that enlarging the mesh size does not affect the 
results presented below.
Two different functionals, namely the Sly4 \cite{Cha98} and the SkM* \cite{Bar82} have been employed in the
mean-field channel.
The specific symmetries imposed in the  {\sl EV8} program only allow for even multipole deformation: quadrupole,
hexadecapole, etc.
The number of nuclei that are found spherical, oblate, prolate, or triaxial with the SkM*
functional are respectively $346$, $59$, $290$ and $54$. The tolerance for the attribution of a given shape is of 0.02 and $1^{\circ}$ respectively for the $\beta_2$ and $\gamma$ value.

The multipole parameters, denoted by $\beta_\lambda$, are rather global quantities that do not allow for a precise analysis
of the local density properties, especially those related to the nuclear surface. To get deeper insight, we have systematically fitted the ground state densities
 using a Thomas-Fermi (TF) shape given by:
 \begin{eqnarray}
\rho({\mathbf r})= \frac{\rho_0}{1+\exp{\left[ \frac{ |{\mathbf r}|-R(\theta,\varphi)}{a(\theta,\varphi)}\right]}}.
\label{eq:rhows}
\end{eqnarray}
Here $({\mathbf r} = |{\mathbf r}|, \theta, \varphi)$ are the standard spherical coordinates.

The radius $R(\theta,\varphi)$ is assumed to take the form:
\begin{eqnarray}
R(\theta, \varphi) = R_0 \left\{1+ \beta_2 Y_{20}(\theta, \varphi)  + \beta_4 Y_{40}(\theta, \varphi)  \right\}
\label{eq:rr}
\end{eqnarray}
where we consider only axially deformed nuclei. As a consequence, the 42 nuclei that are found triaxial are not included in the following analysis. In Eq.~(\ref{eq:rr}) the volume conservation is taken into account in $R_0$.

\begin{figure}[!ht]
	\centering\includegraphics[width=0.5\linewidth]{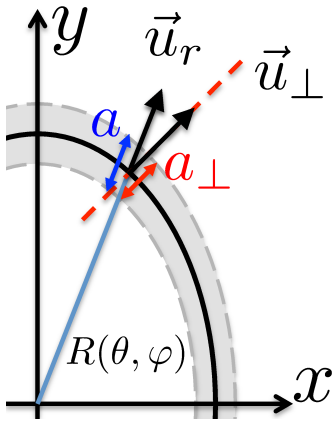}
	\caption{(Color online) Two-dimensional schematic illustration of the difference between
	the diffuseness along the radial axis (along the direction $\vec u_r$) and along the axis perpendicular
	to the isodensity $\rho = \rho_0/2$ (along the direction $\vec u_\perp$).}
	\label{fig:schem}
\end{figure}

The parametrization that should be taken for the diffuseness parameter is less clear. In particular, there is a subtle
aspect related to the fact that $a(\theta, \varphi)$ corresponds to the diffuseness along the radial axis, that is different
from the diffuseness perpendicular to the $\rho(| {\mathbf r}| = R(\theta,\varphi))= \rho_0/2$ isodensity surface.  The latter diffuseness is
denoted by $a_\perp(\theta, \varphi)$ below.  To illustrate this point, a schematic two-dimensional picture is given in Fig.
\ref{fig:schem}.

In spherical systems, the two diffuseness are identical. However, due to the deformation, even if $a_\perp$  is constant,
the radial diffuseness becomes angles dependent.  This has been briefly discussed in Sect. 4 of Ref. \cite{Boh97}.  In this case,
assuming that $a_\perp(\theta, \varphi)= a_0$, the leading correction to the radial diffuseness is given by:
\begin{eqnarray}
a(\theta,\varphi) &\simeq& a_0 \left(1+ \frac{1}{2} \left| \vec \nabla R(\theta,\varphi) |_{r=R(\theta,\varphi)} \right|^2 \right) .
\label{eq:simpledif}
\end{eqnarray}
Such parametrization has two major drawbacks. First, it is only valid at small deformation, which is not always
the case for the considered nuclei. Second, the assumption of a constant $a_\perp$ turns out to be wrong in practice.
For instance, in the situation shown in Fig.  \ref{fig:schem}, a different $a_\perp$ is observed along the dilated ($y$-axis) and  the compressed axis ($x$-axis). This effect cannot be described by the simple expression (\ref{eq:simpledif}), due to the fact that
$\vec \nabla R(\theta,\varphi) =0$ along these two axis.

To overcome these two limitations, we consider a more general form of the diffuseness:
\begin{eqnarray}
a(\theta,\varphi) &=& a_\perp(\theta,\varphi) \sqrt{1+ \left| \vec \nabla R(\theta,\varphi) |_{r=R(\theta,\varphi)} \right|^2 } \label{eq:aa},
\label{eq:compdif}
\end{eqnarray}
that can be easily obtained noting that:
\begin{eqnarray}
\vec u_r.\vec u_{\perp} &=& \frac{1}{\sqrt{1+ \left| \vec \nabla R(\theta,\varphi) |_{r=R(\theta,\varphi)} \right|^2 }} \nonumber
\end{eqnarray}
where the two normalized vectors $\vec u_r$ and $\vec u_{\perp} $ are displayed in Fig. \ref{fig:schem}.
Obviously, this parametrization identifies with Eq. (\ref{eq:simpledif}) for $a_\perp(\theta,\varphi) = a_0$ and
in the leading order of $\left| \vec \nabla R(\theta,\varphi) |_{r=R(\theta,\varphi)}\right|$.

Effects beyond the pure geometric ones are included in the angular dependence of $a_\perp(\theta,\varphi)$.
Here, we assume that this diffuseness can be expressed similarly as in Eq. (\ref{eq:rr}) with:
\begin{eqnarray}
 a_\perp(\theta,\varphi) &=& a_0 \left(1+\tilde \beta_2 Y_{20}(\theta)  + \tilde \beta_4 Y_{40}(\theta) \right).
 \label{eq:atau}
\end{eqnarray}
Note that a similar parametrization has been proposed in ref. \cite{Sax01}.

\section{Results}

For each nucleus, the total local density has been fitted using the  parameters
$\rho_0$, $R_0$, $a_0$, $\beta_2$, $\beta_4$,  $ \tilde \beta_2$, and $\tilde \beta_4$ in Eq.~(\ref{eq:rhows})
with the parametrization (\ref{eq:aa}) together with Eqs. (\ref{eq:rr}) and (\ref{eq:atau}).
The set of parameters obtained for all considered nuclei are provided in the supplemental material
\cite{Supdif}.
In the following, the
terminology "polarization of the nuclear surface" will be employed for systems with non-zero values
of the $ \tilde \beta_2$ and/or $ \tilde \beta_4$ coefficients.
In the present work, only results using the SkM* functional \cite{Bar82} are shown. Note that we also
did the same analysis with the Sly4 \cite{Cha98}  functional (not shown here) leading to
similar conclusions.

\begin{figure}[!ht]
	\centering\includegraphics[width=0.8\linewidth]{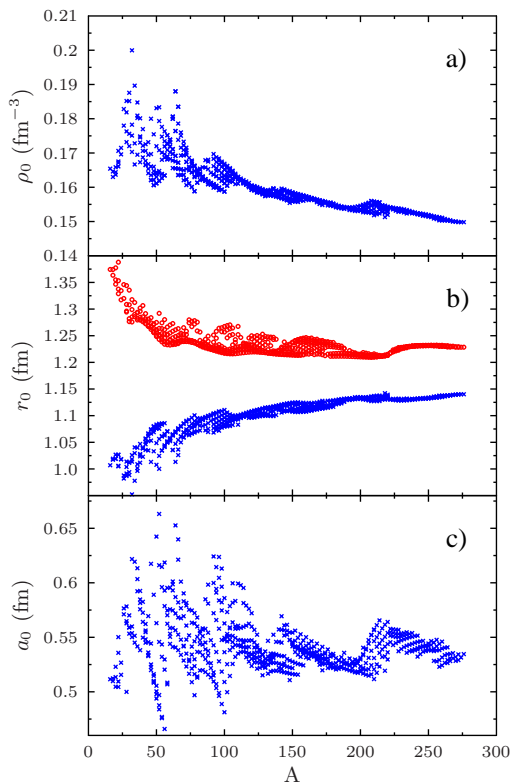}
	\caption{(Color online) Values of the density $\rho_0$ (a),  radius $r_0=R_0/A^{1/3}$ (b) and average diffuseness $a_0$ (c).
	In panel (b), the equivalent sharp radius defined as $r_0 = \sqrt{5/3 \langle r^2\rangle}/A^{1/3}$ is also shown (red open circles)
	for comparison.}
	\label{fig:rra_skms}
\end{figure}

The $\rho_0$, $R_0$, $a_0$ parameters obtained for the SkM*  are respectively shown in Fig. \ref{fig:rra_skms}
for masses $A$ ranging from 16 to 276. The corresponding deformations parameters are shown in Fig. \ref{fig:betaskms}.
Only nuclei that are found to be deformed in {\sl EV8} are shown in these figures.
The central density and extension of nuclei \cite{Ner95,Bro84,Sca13} have been extensively discussed in the literature for instance to study the neutron
skin thickness \cite{Taj01,Sar07} and we only gives here those observables as reference.

The large value of the central density for low masses are due to shell structure effect: the filling of the $s_{1/2}$ level increases the central density in nuclei.
For large masses, this effect is less important. However the average density tends to decrease due to the asymmetry
term proportional to  $\delta=\frac{N-Z}{A}$  that becomes more effective
as the mass increases along the beta stability line \cite{Kho96}.

\subsection{Diffuseness polarization}

Our main focus here is the surface diffuseness. In Fig.~\ref{fig:rra_skms} (c),  large finite size effects are uncovered
in the fluctuations of $a_0$. This is clearly seen in the fluctuations observed for masses $A<150$. The finite size effects are stronger
than in $\rho_0$ and $R_0$, with non-vanishing fluctuations around the mean value $\overline a_0 \simeq 0.55$ even for larger mass. In addition, structures are clearly seen that stems
from the appearance of magic numbers.
 \begin{figure}[!ht]
	\centering\includegraphics[width=0.8\linewidth]{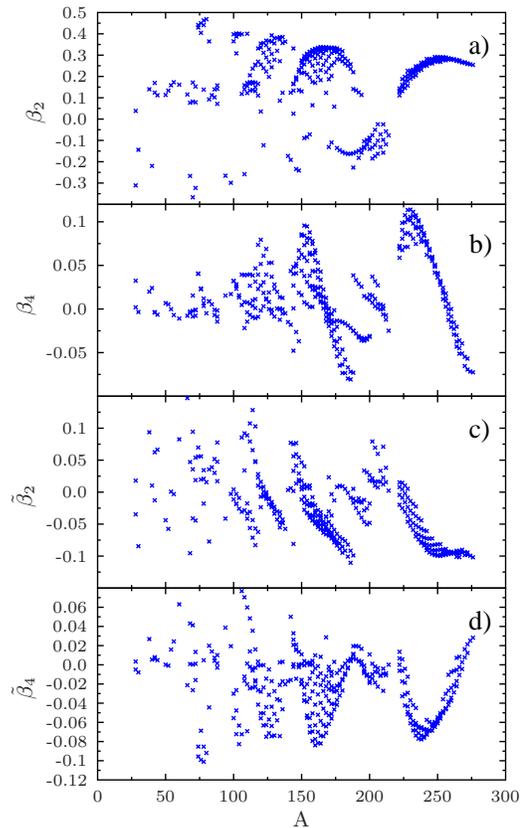}
	\caption{(Color online) Values of the $\beta_2$ (a), $\beta_4$ (b), $\tilde \beta_2$ (c), $\tilde \beta_4$ (d), deduced
	from the fit with the SkM* functional}
	\label{fig:betaskms}
\end{figure}
Figure~\ref{fig:betaskms}, which presents different deformation parameters, carries the main message of the
present work. When the system is deformed, non-trivial distortion of the nuclear surface occurs. This distortion is associated with non-zero values of the $\tilde \beta_2$ and $\tilde \beta_4$ parameters that is called hereafter polarization of the nuclear surface.

\begin{figure}[!ht]
	\centering\includegraphics[width=\linewidth]{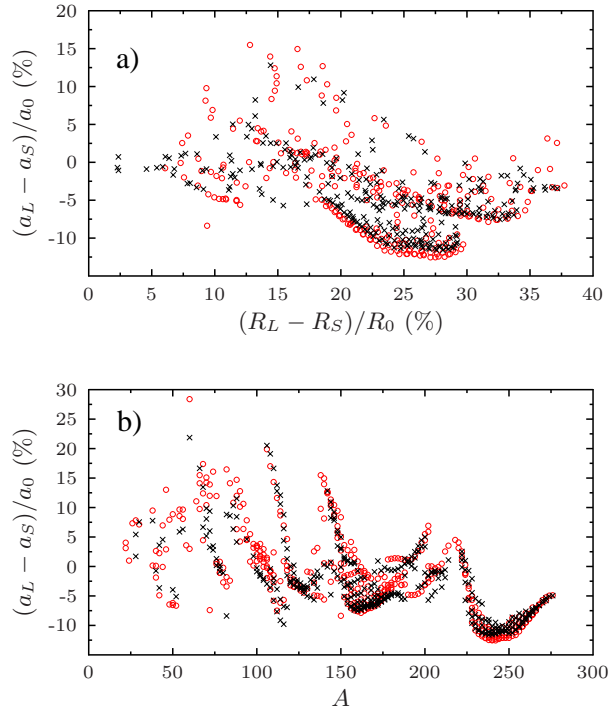}
	\caption{(Color online) Top: $(a_L- a_S)/a_0$ as a function of $(R_L -R_S)/R_0$ for mass $A>120$.
	Bottom: $(a_L- a_S)/a_0$ as a function of mass in $\%$. The black cross and red open circles
	correspond respectively to results obtained with the SkM* and Sly4 functional.}
	\label{fig:aarr}
\end{figure}

For light systems, due to large fluctuations in the deformation parameter, a systematic
tendency of the surface distortion can hardly be uncovered.  For mass $A>120$, fluctuations
are strongly suppressed.  In particular in the heavy mass region, we clearly see that an increase
of $\beta_2$ (with $\beta_2 >0$) leads to increase of $|\tilde \beta_2|$ ($\tilde \beta_2<0$. Denoting, by $R_L$ (resp. $a_L$)
and $R_S$ (resp. $a_S$) the radius (resp. the diffuseness) along the elongated and compressed  axis,
the two quantities $(R_L -R_S)$ and $(a_L- a_S)$ are strongly
anti correlated (see Fig. \ref{fig:aarr}) in this mass region.
Note that, with the present deformed Fermi densities, we have
\begin{eqnarray}
\frac{(R_L- R_S)}{R_0} & = & \frac{3}{4} \sqrt{\frac{5}{\pi}} \beta_2 + \frac{15}{16\sqrt{\pi}}\beta_4,
\end{eqnarray}
with a similar expression for $(a_L- a_S)/a_0$ except that the $\beta$'s are replaced by the $\tilde \beta$'s. 
In Fig.~\ref{fig:aarr}, the
results obtained using the Sly4 functional are also shown, demonstrating that changing the functional
leads qualitatively and quantitatively to the same effect.

\begin{figure}[!ht]
	\centering\includegraphics[width=1.0\linewidth]{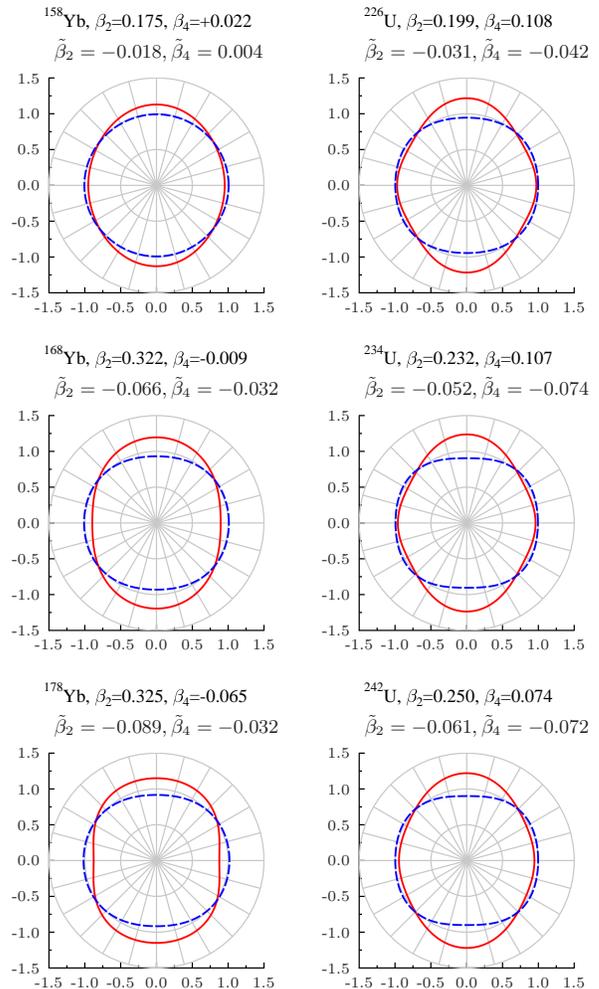}
	\caption{(Color online) Illustration of the shape and diffuseness distortion for selected Yb and U
	isotopes. The quantities  $R(\theta,\varphi=0)/R_0$  (red solid line) and $a(\theta,\varphi=0)/a_0$ (blue long dashed line) are shown
	as a function of $\theta$ in spherical coordinate representation.
	}
	\label{fig:shape}
\end{figure}

Therefore, in heavy systems, one can systematically observe the following phenomena: as the system becomes
more and more deformed, its diffuseness along the elongated axis becomes smaller in favor of an increase of
the diffuseness along the compressed axis. This effect is illustrated in Fig. \ref{fig:shape} along the isotopic chain of 
Yb and U isotopes.

Such distortion of the nuclear surface is a highly non-trial effect that stems from a complex mixing
of the volume, surface and Coulomb field entering in the EDF in the presence of pairing. Surprisingly enough, this effect
does not seem to be negligible and can lead to an overall fluctuations of 10 $\%$ of the diffuseness along the isodensity
contour (Fig.~\ref{fig:aarr} (b)).

\subsection{ Neutron-proton effects}

Above, we have concentrated our attention on the total density where possible isospin effects are averaged out.
Possible $N/Z$ dependence of the nuclear density profiles can be analyzed by considering separately the neutron and proton densities, denoted
by $\rho_\tau ({\mathbf r})$, with $\tau=n$, $p$, respectively for neutron and proton.
Then, the same fitting procedure as above can be used, leading to two sets of parameters.
The set of parameters obtained for all considered nuclei separately for neutrons and protons
are provided in the supplemental material \cite{Supdif}.

\subsubsection{Nuclear densities and radius}

\begin{figure}[!ht]
	\centering\includegraphics[width=\linewidth]{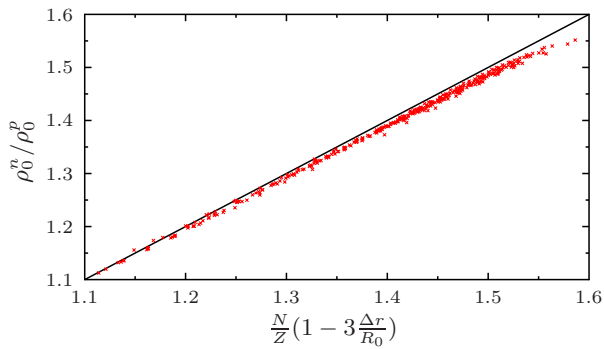}
	\caption{(Color online) $\rho_{0}^n/\rho^p_{0}$ as a function of 
$\frac{N}{Z} \left( 1 - 3 \frac{\Delta r}{R_0} \right)$. The $x=y$ line is added
	as a guidance. Only nuclei with $A>120$ are shown. }
	\label{fig:rhonrhopcor}
\end{figure}

As has been widely studied, when the $N/Z$ ratio changes, we do anticipate specific
behavior of the neutron skin thickness, that can be related to the symmetry energy \cite{Vin12,Che05,Hor01}.
We give here specific aspects related  to the use of EDF in combination with the Thomas-Fermi shape analysis of the density profile.
Let us assume in first approximation that nuclei can be considered as non-deformed nuclei and that the two proton
and neutron Fermi liquids present both an equivalent sharp radius (ESR), denoted by $R'_n$
and $R'_p$. The average densities will verify $\rho^n_0/\rho^p_0 = N/Z (R'_n/R'_p)^3$, leading to
\begin{eqnarray}
\frac{\rho^n_0}{\rho^p_0} &\simeq& \frac{N}{Z} \left( 1 - 3 \frac{\Delta r_{np}}{R'_0} \right) \label{eq:rnrp},
\end{eqnarray}
where $R'_0$ is the equivalent sharp radius of the total (neutron+proton) droplet while $\Delta r_{np}= R'_n -R'_p$
is the neutron skin thickness. Approximate expression of the ESR in terms of the TF shape parameters is obtained
using  Eq.  (2.64) of Ref.~\cite{Boh97-1}:
\begin{eqnarray}
\frac{\rho^n_0}{\rho^p_0} &\simeq& \frac{N}{Z} \left( 1 - 3 \frac{\Delta r}{R_0} \right)
\left(
\frac{1 + \pi^2\frac{a^2_{p}}{R^2_p} }
{1 + \pi^2 \frac{a^2_{n}}{R^2_n} }
 \right)
 \label{eq:rra}
\end{eqnarray}
where $\Delta r  =  R_n - R_p$.
Equation (\ref{eq:rra}) provides an approximate
analytical expression of the correlation between the density, radius and diffuseness
induced by the neutron and proton number constraints.  Fig. \ref{fig:rhonrhopcor}  illustrates that this relation is perfectly
fulfilled and that the contributions of the surface and deformation on average densities are almost negligible.

 \begin{figure}[!ht]
	\centering\includegraphics[width=1\linewidth]{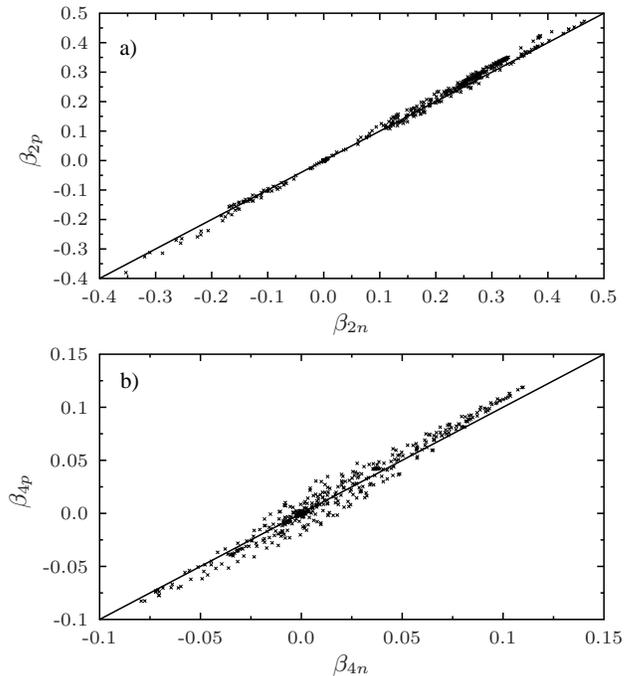}
	\caption{(Color online) Correlation between the neutron and proton deformation parameter: (a) $\beta_{2n}$ vs $\beta_{2p}$,
	(b) $\beta_{4n}$ vs $\beta_{4p}$. The line $x=y$ are again to guide the eyes. }
	\label{fig:b2b4np}
\end{figure}
\begin{figure}[!ht]
	\centering\includegraphics[width=0.8\linewidth]{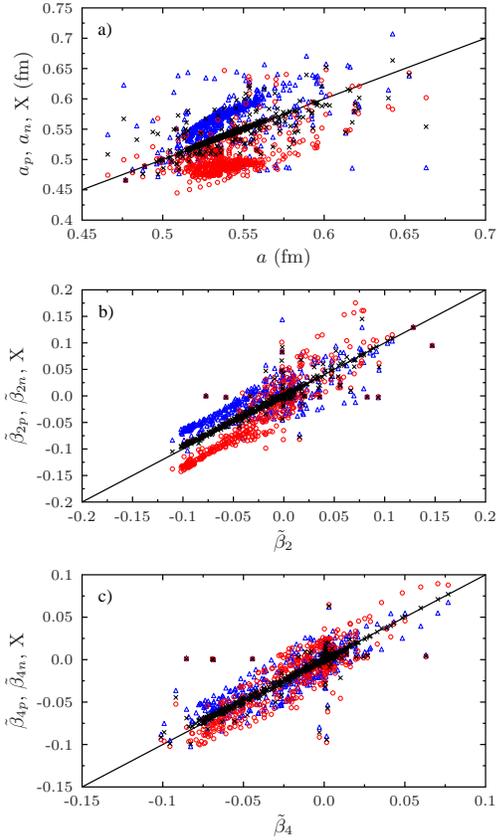}
	\caption{(Color online) Correlation between the neutron or proton parameters associated to the surface diffuseness:
	(a) $a_{n}$ (blue triangles) and $a_{p}$ (red circles) as a function of $a_0$,
	(b) $\tilde \beta_{2n}$  (blue triangles)  and $\tilde \beta_{2p}$ (red circles) as a function of $\tilde \beta_2$,
	(c) $\tilde \beta_{4n}$  (blue triangles)  and $\tilde \beta_{4p}$ (red circles) as a function of $\tilde \beta_4$.
	For each quantity, the weighted averages $X=(N/A) X_{n} + (Z/A) X_{p}$ (see text) are also shown by
black crosses.
	}
	\label{fig:a0beta2beta4np}
\end{figure}

\subsubsection{$N/Z$ ratio and deformation}

To trace back possible isospin dependence of the deformation, we show in panel (a) (resp. panel (b)) of Fig. \ref{fig:b2b4np}, the
correlation between $\beta_{2n}$ (resp. $\beta_{4n}$) and the $\beta_{2p}$ (resp. $\beta_{4p}$).
In addition, possible interplays between isospin, diffuseness and deformation are illustrated
in Fig. \ref{fig:a0beta2beta4np}. From these two figures, one can draw the following conclusions:
(i) The shape deformations of nuclei appear to be almost independent on the fact that protons, neutrons
  or both are considered, i.e. $\beta_{xn} \simeq \beta_{xp} \simeq \beta_{x}$ with $x=2$ or $4$.
(ii) The diffuseness properties depend explicitly of the neutron/proton nature of the considered fluid.
Not surprisingly due to the absence of the Coulomb field, neutrons are generally more diffuse than protons.
In addition,  they present an enhanced deformation of the diffuseness, especially in the quadrupole
parameter $\tilde \beta_{2\tau}$, the $\tilde \beta_{4\tau}$ factors being globally the same.  (iii) Denoting by $X_\tau$ one of the parameters ($a_\tau$, $\beta_{2\tau}$ or $\beta_{4\tau}$)  associated
  to the diffuseness and by $X$ the equivalent quantity obtained by fitting the total density, to a good approximation, we found that:
  \begin{eqnarray}
X & \simeq & \left(\frac{N}{A}\right) X_n +  \left(\frac{Z}{A}\right) X_p.
\end{eqnarray}
 \begin{figure}[!ht]
       \centering\includegraphics[width=1.0\linewidth]{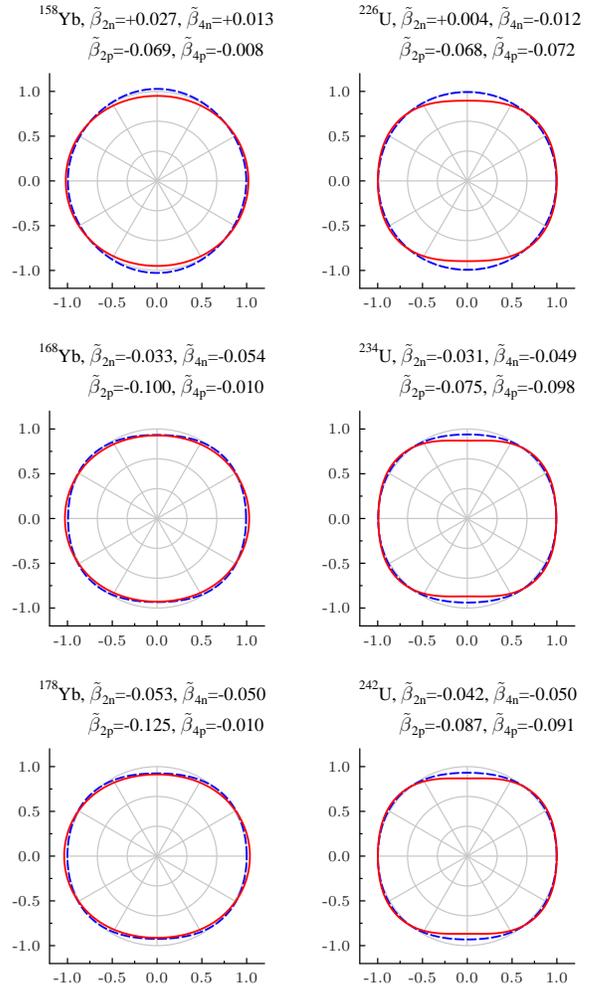}
	\caption{(Color online) Illustration of the neutron and proton
	diffuseness distortion for selected Yb and U
	isotopes. The quantities  $a_n(\theta,\varphi=0)/a_{n}$ (blue long dashed line) and $a_p(\theta,\varphi=0)/a_{p}$ 
are shown
	as a function of $\theta$ (red solid line) in spherical coordinate representation.}
	\label{fig:shape2}
\end{figure}
As a quantitative illustration of the isospin dependence of the diffuseness as well its possible distortion
when the system is deformed, the angular dependences of the proton and neutron diffuseness for selected Yb and
U isotopes are shown in Fig. \ref{fig:shape2}.
 \begin{figure}[!ht]
       \centering\includegraphics[width=1.0\linewidth]{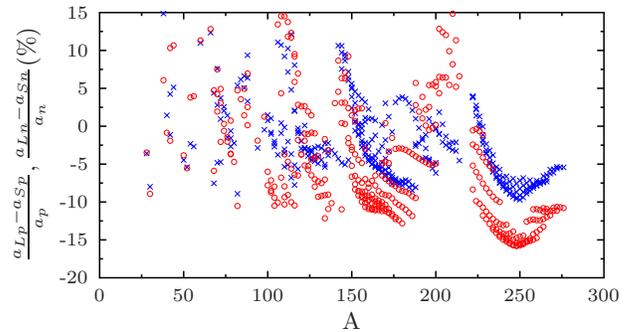}
	\caption{(Color online) $(a_{L\tau}- a_{S\tau})/a_{\tau}$ as a function of mass in $\%$ for proton (red open circles)
	and neutrons (blue cross). }
	\label{fig:anpvsa}
\end{figure}

To quantify systematically the polarization of the neutron/proton
density profile, the quantities  $(a_{L\tau}- a_{S\tau})/a_\tau$ are shown
as a function of mass for all deformed nuclei considered in this work.
We see that the relative polarization is bigger for proton compared to neutrons.
Note however that the absolute value of $a_n$ is in general larger than the one of $a_p$. It should
also be noted that the charge density of heavy-systems tends to develop a hole at the
center of the nucleus. In that case, the TF assumption for the density becomes
less accurate.

\subsection{Pairing effect}

\begin{figure}[!ht]
       \centering\includegraphics[width=1.0\linewidth]{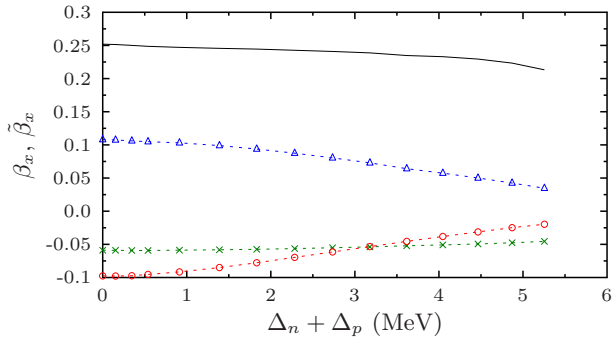}
	\caption{(Color online) Deformation parameters in $^{238}$U as a function of the proton+neutron pairing
	gap: $\beta_2$ (black solid line), $\beta_4$ (blue open triangles), $\tilde \beta_2$ (green crosses) and $\tilde \beta_4 $ (red open circles). }
	\label{fig:pair_effect}
\end{figure}

The pairing correlation is known to play a role in the deformation. A general effect, is that the shape of the nuclei tends to spherical
symmetry when the pairing gap increases. Here, we want to see if a similar effect arises for the polarisation of the diffuseness. To analyze the possible effect of pairing, we change the pairing interaction with different values from $V_0^{nn}=0$ to 2200 MeV$\;$fm$^3$ with a proton-proton interaction proportional to the neutron-neutron interaction $V_0^{pp}=1.125 V_0^{nn}$.
As the pairing interaction increases, the neutron+proton pairing gap increases from 0 to 5 MeV. Fig. \ref{fig:pair_effect} shows a systematic behavior that reduces the values of the quadrupole and hexadecapole deformations of $^{238}$U when the pairing gap increases. The same behavior is found for the $\tilde \beta_2$ and $\tilde \beta_4 $ parameters and a direct correlation can be establish between the $\beta_x$ and the corresponding $\tilde \beta_x$
confirming the conclusions already drawn from Fig. \ref{fig:betaskms}.
It should be noted that the results present in the Fig. \ref{fig:pair_effect} are almost insensitive to the pairing interaction type (surface, mixed or volume).

\section{Illustration of possible surface polarization effects on nuclear dynamics}

In the previous section, we have shown that the onset of deformation induces systematically a polarization
of the nuclear surface diffuseness that depends on the isospin. Such a polarization has sometimes been
suspected to affect nuclear dynamics properties. For instance, in Ref.~\cite{Pen09}, the surface diffuseness
polarization has been proposed as a source of reduction of the GDR strength in the low-lying energy sector.

Another anticipated  effects is a possible modification of the fusion barrier due to the change
of diffuseness. This aspect might be particularly crucial for reactions involving very heavy systems
generally used in the super-heavy element quest. In this case, even a slight change in the physics of the path towards
fusion can lead to large effects in the evaluated capture cross-sections.

\begin{figure}[h]
\centering
\includegraphics[width=0.9 \linewidth,clip]{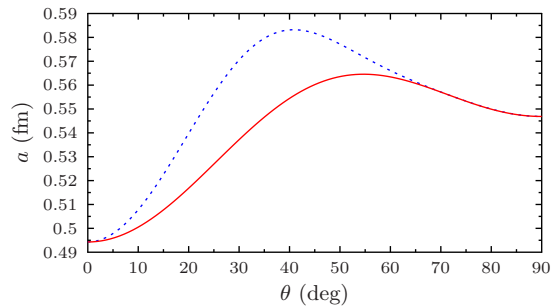}
\caption{(Color online) Evolution of  the radial (blue dotted line) and perpendicular (red solid line)
diffuseness in $^{238}$U are shown as a function of $\theta$ for $\varphi=0$. }
\label{fig:athetaU}
\end{figure}

Consistently with the double-folding approach and/or proximity potential approach, a change
in the nuclear diffuseness is anticipated to influence the
Coulomb barrier properties. Here, we give an illustration for the reaction $^{40}$Ca+$^{238}$U. The uranium nuclei have been shown in Figs. \ref{fig:shape} and \ref{fig:shape2}
to present significant deformation in their ground states inducing diffuseness polarization.
In Fig. \ref{fig:athetaU}, the quantitative dependence of the radial and perpendicular diffuseness  for $\varphi=0$
are shown as a function of $\theta$ for $^{238}$U. The diffuseness shown here is the one associated
 to the total density.
Due to the non-negligible deformation ($\beta_2=0.244$, $\beta_4=0.094$) found with SkM*,
a significant polarization of the diffuseness is observed.

To estimate the impact of diffuseness change on nuclear dynamics, the nucleus-nucleus
interaction potential $V(R)$ has been estimated using three different values of the
diffuseness parameter
using the procedure presented in Refs.~\cite{Adamian,Sar10,Sar11}.
For the nuclear part of the nucleus-nucleus
potential, the double-folding formalism with
density-dependent effective
nucleon-nucleon interaction is used.
Within this approach many heavy-ion capture
reactions with stable and radioactive beams
at energies above and well below the Coulomb barrier have been
successfully described~\cite{Sar12}.
The deformations of colliding nuclei are taken into account in this approaching phase.
Here, we take into account the quadrupole and hexadecapole deformation
of the $^{238}$U found in the {\sl EV8} code. For the spherical nucleus,
a diffuseness $a=0.59$ fm is used while for the deformed $^{238}$U, the three values
$a=0.494$, 0.5504, and 0.5832 fm. These values correspond respectively to the
minimal, average and maximal values of the radial diffuseness shown in Fig.~\ref{fig:athetaU}.
\begin{figure}[h]
\centering
\includegraphics[width=7cm,clip]{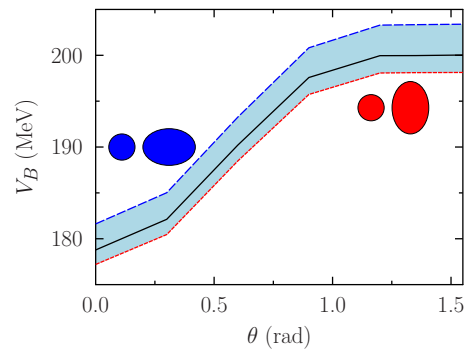}
\caption{The calculated dependencies of the Coulomb-barrier heights $V_B$ on the
 mutual orientation of colliding nuclei $^{40}$Ca and $^{238}$U.  The three curves
 have been obtained using a diffuseness parameter of $a=0.494$ (blue long dashed line), 0.5504 (black solid line) and
 0.5832 fm (red short dashed line) and deformation parameter for the $^{238}$U are  $\beta_2=0.244$, $\beta_4=0.094$.
 The filled area is shown to underlines the possible change in the fusion barrier.}
\label{fig:VbbCaU}
\end{figure}
We see from this figure, that the change of diffuseness can induce an increase
or a decrease of the fusion barrier and will ultimately modify the capture
cross-section. An increase of the barrier is anticipated for zero relative angle
due to the lower diffuseness along the main axis of deformation, while at $\theta= \pi/2$
the barrier will be reduced. The influence of this effect on the nucleus-nucleus interaction
deserves further investigations.
Note that the saturation at large $\theta$ is due
to the interplay between the nuclear and Coulomb interactions. 

\section{Conclusion}

In the present work, a detailed analysis of the possible diffuseness polarization was made for deformed nuclei. 
It is observed, especially in medium- and heavy-mass nuclei that
the diffuseness along the main deformation axis tends to reduce as the nuclei becomes more and more deformed, while
the opposite is seen along other axis. Such deformation seems to be a generic effects that is predicted by the Skyrme
energy density functional independently on the effective interaction that is used.
It was also shown that the polarization of the nuclear surface for neutrons slightly differs from the one for protons and that pairing correlation have a strong influence on the surface deformation and skin polarization.
The polarization of the nuclear surface can directly affect some aspects related to nuclear dynamics. Among them, we anticipate
specific features in collective motion built on deformed nuclei. Besides, small amplitude vibrations, we quantitatively
 illustrated the possible
modification of the fusion barrier  for the $^{40}$Ca+$^{238}$U reaction.

\end{document}